\documentclass[conference]{IEEEtran}

\usepackage{fancyhdr}

\IEEEoverridecommandlockouts
\usepackage{cite}
\usepackage{multirow}
\usepackage{amsmath,amssymb,amsfonts}
\usepackage{algorithmic}
\usepackage{graphicx}
\usepackage{textcomp}
\usepackage{xcolor}
\def\BibTeX{{\rm B\kern-.05em{\sc i\kern-.025em b}\kern-.08em
    T\kern-.1667em\lower.7ex\hbox{E}\kern-.125emX}}

\usepackage[T1]{fontenc}
\usepackage[utf8]{inputenc}
\usepackage[hidelinks]{hyperref}
\usepackage{booktabs}
\graphicspath{{build/img/}{img/}}

\renewcommand{\itemautorefname}{}

\makeatletter
\renewcommand{\itemautorefname}{\@gobble}
\makeatother

\begin{document}

\title{
  Collecting Infrastructure-Level Measures for Energy Analysis in Modern Computing Infrastructures
  }

 \author{
   \IEEEauthorblockN{Isac Pasianotto\textsuperscript{$\dagger$}}\
    \IEEEauthorblockA{\textit{University of Trieste} \\
    \textit{Area Science Park} \\
    Trieste, Italy \\
    \href{mailto:isac.pasianotto@phd.units.it}{isac.pasianotto@phd.units.it} \\
    \href{https://orcid.org/0009-0000-9759-5706}{ORCID: 0009-0000-9759-5706}
    }
    \and
    \IEEEauthorblockN{Stefano Cozzini}
    \IEEEauthorblockA{\textit{AREA Science Park} \\
    Trieste, Italy \\
    \href{mailto:stefano.cozzini@areasciencepark.it}{stefano.cozzini@areasciencepark.it} \\
    \href{https://orcid.org/0000-0001-6049-5242}{ORCID: 0000-0001-6049-5242}
    }
    \and
    \IEEEauthorblockN{Niccolo Tosato\textsuperscript{$\dagger$}}
    \IEEEauthorblockA{\textit{University of Trieste} \\
    \textit{Area Science Park} \\
    Trieste, Italy \\
    \href{mailto:niccolo.tosato@phd.units.it}{niccolo.tosato@phd.units.it} \\
    \href{https://orcid.org/0009-0005-4996-9933}{ORCID: 0009-0005-4996-9933}
    }

    \thanks{\textsuperscript{$\dagger$}Both authors equally contributed to this work.}
}

\maketitle

\thispagestyle{fancy}
\fancyhf{}
\lfoot{\parbox[c]{\textwidth}{\footnotesize
  \copyright~2026 IEEE. Personal use of this material is permitted. Permission
  from IEEE must be obtained for all other uses, in any current or future media,
  including reprinting/republishing this material for advertising or promotional
  purposes, creating new collective works, for resale or redistribution to
  servers or lists, or reuse of any copyrighted component of this work in other
  works.\\
  Accepted version. Published in: SC26 Workshops, November 15--20, 2026,
  Chicago, Illinois, USA. DOI: \href{https://doi.ieeecomputersociety.org/10.1109/SCW72768.2026.00200}{10.1109/SCW72768.2026.00200}}}
\renewcommand{\headrulewidth}{0pt}
\renewcommand{\footrulewidth}{0pt}

\begin{abstract}

Modern HPC facilities are complex environments in which all components are tightly coupled; consequently, precisely modeling and accounting for energy consumption requires approaches that go beyond instrumenting individual compute nodes.
Aiming for a cluster-wide vision, we designed a scalable data collection pipeline that aggregates fine-grained measurements from multiple subsystems---spanning component loads, software and hardware counters, sensors and energy---ensuring consistent sampling frequencies and synchronized data acquisition across the infrastructure.
We present a dataset of comprehensive metrics collected at 1 Hz over one month from a production research data center supporting heterogeneous workloads, including HPC, Kubernetes services, and distributed storage.
To illustrate what this approach makes possible in practice, we analyze two case studies---distributed I/O and large-language-model inference---showing how cluster-wide measurement reveals system-level dynamics missed by localized instrumentation.
The proposed dataset and collection methodology provide a foundation for future work on predictive modeling, energy-aware resource management, and more efficient computing infrastructures.
\end{abstract}


\section{Introduction}
\label{section:intro}

Despite a massive surge in computational demand, global data-center energy consumption remained effectively flat between 2010 and 2022, at approximately 200~TWh, thanks to substantial gains in hardware and infrastructure efficiency---the latter commonly tracked through the Power Usage Effectiveness (PUE) metric 
\cite{IEA2020DataCentreEnergy,recalibrating}.
The explosive growth of artificial intelligence and cloud services has now broken this trend: global data-center demand reached 400~TWh in 2024---a $2\times$ increase over 2020, about 1.5\% of worldwide electricity consumption---and is projected to reach 800~TWh by 2030 \cite{IEA2025EnergyAI}.

While hardware efficiency is aggressively optimized by vendors, the energy behavior of software and the infrastructure has received considerably less attention. 
Energy-aware algorithms, schedulers, and workflows promise savings without new hardware, but every such optimization requires precise energy measurements and profiling.

In a modern facility, compute nodes are only one part of the overall energy footprint.
Many necessary components, such as distributed file systems, monitoring infrastructure, and network components draw power continuously.
As a result, energy consumption is distributed across tightly coupled subsystems that cannot be meaningfully analyzed in isolation, and lastly, the IT equipment necessary to run the infrastructure itself is not always taken into account.

A further, more subtle obstacle is the nature of the data that facilities typically collect.
Operational monitoring stacks are designed for dashboards and alerting: they scrape metrics on a best-effort basis, at sparse or adaptive intervals, downsample aggressively as data ages, and treat occasional gaps as acceptable.
Data of this kind cannot support quantitative energy analysis, where power series must be integrated over time, cross-correlated across subsystems, and joined at fine granularity with application events.
What is needed instead is measurement: fixed-frequency sampling, synchronized timestamps across the whole infrastructure, lossless collection, and validated completeness.

In this work we report the design, deployment, and validation of a measurement pipeline covering an entire production research data center---compute, storage, network, cloud services, and facility-level power---and we release the resulting dataset. 
Our contributions are twofold:

\begin{enumerate}
    \item \textbf{A reproducible measurement pipeline.} We describe a collection architecture assembled exclusively from open-source components.
      The pipeline samples at a fixed frequency with cross-node timestamp alignment, and we quantify its collection overhead, its vertical and horizontal scaling headroom (\autoref{sec:pipeline}).
      Any center can replicate the setup with commodity components and produce its own dataset; we distribute the deployment configurations alongside the data.
    \item \textbf{A public, high-quality dataset.} We release 28 days of continuous 1~Hz telemetry measuring from a production facility running heterogeneous workloads (HPC jobs, Kubernetes services, distributed storage): more than $2.4\times10^{10}$ rows across 11 telemetry tables, complemented by the Slurm job database table and by the token-level billing records of a production LLM inference service (\autoref{sec:dataset}).
      Unlike existing job-centric datasets, ours consists of raw, non-aggregated time series spanning every infrastructure layer, whose completeness (i.e., absence of gaps) and temporal alignment (i.e., timestamp synchronization across sources) are explicitly assessed.
\end{enumerate}

To demonstrate what this combination enables, we present two case studies (\autoref{sec:cases}).
The first quantifies the energy efficacy of distributed I/O on a production Ceph file system, relating device-level activity on the storage servers to the bandwidth actually delivered to applications, and explaining the I/O amplification in between.
The second provides a per-component energy account of LLM inference, showing that GPUs represent less than half of the facility-level running cost and that reporting based on GPU counters alone underestimates true consumption by $2.4\times$--$4.6\times$ once whole-node consumption and the actual PUE correction is taken into account.

\section{Related Work}
\label{sec:related}

\paragraph{Measuring energy in HPC systems}
Methodologies for energy measurement have evolved from processor-local interfaces towards system-level instrumentation. Early workload characterizations relied on Running Average Power Limit (RAPL) counters for the CPU and memory domains \cite{rapl}; later work incorporated external sensors, Baseboard Management Controllers (BMCs) queried through the Intelligent Platform Management Interface (IPMI), and rack-level Power Distribution Units (PDUs) \cite{C6EnPLS,exploringIO,practicalPower}.
Each interface covers a different slice of the stack, with different accuracy and latency; a facility-wide picture requires combining all of them, which is precisely what our pipeline does.

\paragraph{Continuous monitoring frameworks}

Several HPC centers operate large-scale telemetry infrastructures. LDMS demonstrated lightweight, high-frequency metric collection on very large systems \cite{ldms}, though its scope remains system-level telemetry and does not target energy attribution; Examon introduced an MQTT-based holistic monitoring architecture and powered the datasets later released for the Marconi100 system \cite{examon}; DCDB integrates facility- and application-level sensor data in a single modular framework \cite{dcdb}. More recently, storage-side monitoring with coarse (30--60~s) scrape intervals has been used to attribute I/O energy on the Lustre parallel file system \cite{exploringIO}.
Our work advances the state of the art along three axes that, to the best of our knowledge, no prior system addresses jointly. First, \emph{coverage}: we unify compute, storage, network, cloud services, and facility power under a single, consistent schema, rather than the single-domain or facility-only views offered by existing frameworks. Second, \emph{measurement discipline}: we enforce a fixed sampling grid with validated alignment and completeness and no downsampling---a rigor that existing monitoring deployments rarely enforce, and one that is essential for reliable, fine-grained energy attribution. Third, \emph{reproducibility}: our stack is built entirely from unmodified, widely packaged open-source components, deliberately chosen so the deployment can replicated without custom software or vendor-specific engineering effort.

\paragraph{Public datasets}
Datasets correlating power with workload features have been released for several production systems. F-DATA covers Fugaku jobs with job-centric features \cite{fdata}; PM100 provides power data for roughly 230K jobs of the Marconi100 system \cite{pm100} and has been used for job-level carbon-efficiency modeling \cite{pm100model}; C6EnPLS targets linear-algebra workloads with physical-sensor readings from a production cluster \cite{C6EnPLS}.
These datasets share three structural limitations: they represent each job as a single row of descriptive statistics (or as coarse averages) rather than as time series; they cover compute nodes only, excluding storage, network, and facility subsystems; and their underlying sampling is sparse.
Our dataset takes the complementary approach: raw 1~Hz series for every component, released together with the job table and service-level accounting records, so that job-centric views can be \emph{derived} from it rather than being the only view available.

\paragraph{Modeling and energy-aware operation}
A rich body of work builds predictive models on such data: job power prediction from scheduler submission data \cite{Predicting}, online power forecasting on Fugaku and Marconi100 \cite{onlinePrediction}, software-based estimation validated against hardware power measurements \cite{practicalPower}, and Slurm plugins for energy-efficient scheduling \cite{automaticEnergy}.
These efforts consistently identify missing observability as a limiting factor, in particular for I/O: associating jobs with the energy consumed on their behalf by parallel file systems remains an open problem, and the profiling literature explicitly calls for server-side file-system statistics \cite{parallelIO}. Recent work attributes Lustre energy to jobs through node-local burst buffers and coarse storage-server monitoring \cite{exploringIO}. Our dataset is designed to serve exactly this line of research, providing the storage-side and network-side series that compute-centric datasets lack.

\paragraph{Energy accounting for LLM inference}
The energy and carbon footprint of LLM inference is attracting growing attention \cite{advocating}, and practical estimation tools such as CodeCarbon model consumption from GPU, CPU, and RAM figures \cite{codecarbon}.
Most published numbers, however, are accelerator-local: they neglect the rest of the serving node (fans, interconnect fabric, storage, idle accelerators held by the same deployment) and the facility overhead. Both are far from negligible, as our second case study quantifies (\autoref{sec:llm}).


\section{Measurement Pipeline}
\label{sec:pipeline}

Time-series analysis for energy modeling routinely fails on data produced by conventional monitoring stacks: missing intervals, uneven sampling, inconsistent frequencies across sources, and misaligned timestamps make integration and cross-source joins unreliable.
We therefore designed the pipeline around six requirements:

\begin{enumerate}
  \renewcommand{\theenumi}{R\arabic{enumi}}
  \renewcommand{\labelenumi}{\theenumi}
  \item \label{req:R1} \emph{Fixed-frequency sampling.} Every observable is polled on a fixed frequency (e.g., 1~Hz), so that power can be integrated into energy and series can be joined without resampling artifacts.
  \item \label{req:R2} \emph{Temporal alignment.} Timestamps must be synchronized across nodes and subsystems. All nodes stamp samples with nanosecond precision using the system clock, which is synchronized infrastructure-wide via the Network Time Protocol (NTP).
  \item \label{req:R3} \emph{Losslessness.} Transient network congestion or database pressure must not create gaps; collected data are never downsampled or expired, unlike typical monitoring retention policies.
  \item \label{req:R4} \emph{Data comprehensiveness} Every powered device is accounted for: either instrumented natively, or---when mission-critical or closed (switches, gateways, storage appliances)---covered by external PDU and energy-meter measurements. 
  \item \label{req:R5} \emph{Non-intrusiveness.} The collection itself must not perturb the workloads being measured. 
  \item \label{req:R6} \emph{Portability and scalability.} The stack must deploy on heterogeneous operating systems without touching system packages, and must scale both vertically and horizontally as nodes are added.
\end{enumerate}

Treating collection as \emph{measurement} rather than monitoring, means enforcing all six requirements, and \autoref{subsec:quality} verifies \autoref{req:R1}--\ref{req:R3} empirically on the production deployment.

\subsection{Node-level collection}
\label{subsec:node-level}

All infrastructure nodes---compute, storage, and service nodes alike---run a Telegraf\footnote{\url{https://github.com/influxdata/telegraf}} agent.
Telegraf is an open-source collection daemon whose plugins bridge system interfaces (e.g., kernel counters, hardware registers, IPMI, ...) to a time-series output; its architecture supports per-plugin fixed polling intervals and timestamp precision (\autoref{req:R1}, \autoref{req:R2}).
Three properties motivated its selection over alternatives.
First, its ecosystem: nearly 250 maintained input plugins cover current server hardware, and several output plugins---e.g. InfluxDB Line Protocol (ILP), Kafka, MQTT, SQL, OpenTelemetry, and Parquet---so the agent layer does not lock the pipeline to any specific database.
Second, its transmission model: agents push their measurements, buffering locally when the database is unreachable (\autoref{req:R3}), and pushes are staggered across nodes to spread network and ingestion load evenly over time (jittering).
Third, its deployment model: the agent ships as a container image, which we run through Podman Quadlet as a systemd service. The same artifact deploys unmodified on every node regardless of the underlying Linux distribution, without installing packages on the hosts (\autoref{req:R5}, \autoref{req:R6}).

\subsection{Facility-level collection}
\label{subsec:facility-level}

Node-local telemetry cannot capture the facility-level overhead of cooling and power delivery (\autoref{req:R4}).
Two additional collectors close this gap.
The first performs an SNMP walk over the rack PDUs, retrieving per-socket power for every connected device, with each socket labeled with the attached equipment.
The second queries the facility energy meters through their REST API; these meters report active and reactive power, voltages, currents, and cumulative energy for the electrical lines that feed the datacenter as a whole, and provide the measurements from which the facility PUE is computed.

\subsection{The Ingestion Tier and Long-Term Storage}
\label{subsec:database}

Agents write into a time-series database (TSDB) whose sole role is to absorb the telemetry stream and retain it until the next export cycle. Since it serves no interactive queries or dashboards, ingestion throughput is the primary selection criterion, leading to the adoption of QuestDB (\autoref{subsec:scalability}) in our case.
To maximize throughput, the TSDB stores data uncompressed, making long-term retention within the ingestion tier impractical---our deployment produced approximately 2.3~TB per month.
Data are therefore periodically exported to Parquet files using \texttt{lz4\_raw} compression, reducing the monthly footprint to approximately 366~GB, a $6.28\times$ compression ratio.
Each table is partitioned into 24-hour Parquet chunks, a granularity that keeps individual files manageable while preserving efficient access for time-bounded analyses.
The resulting Parquet objects are stored in an S3 bucket, over which Apache Iceberg tables are defined and exposed via an Apache Polaris metadata catalog, providing predicate pushdown and metadata-based file pruning while decoupling the archival format from the ingestion technology.
Both the QuestDB-based ingestion layer and the Apache Arrow-based long-term storage layer are distributed as containers, natively deployable on Kubernetes (\autoref{req:R6}).

\subsection{Scalability}
\label{subsec:scalability}

As mentioned in the previous section, the most relevant criterion in TSDB selection is ingestion throughput, since the ingestion layer can easily become a bottleneck of the overall architecture.

\begin{figure}[htbp]
    \centering
 		\includegraphics[width=\columnwidth]{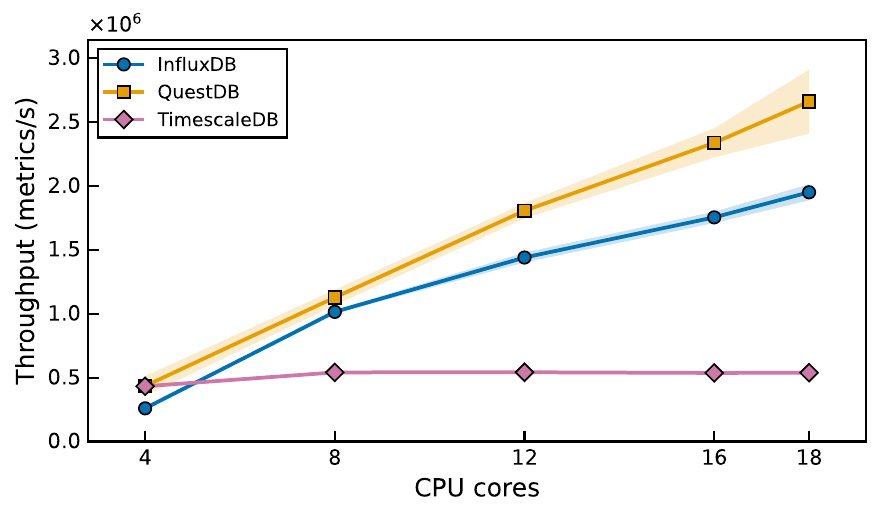}
    \caption{
        TSDB ingestion benchmark results: metrics inserted per second for assessed TSDBs, with 64~GB of RAM and an increasing number of cores. Each configuration was run 10 times.
    }
        \label{fig:tsdb}
\end{figure}

We benchmarked three candidates\footnote{
  We restricted the evaluation exclusively to open-source alternatives.
  }---QuestDB, InfluxDB~v2, and TimescaleDB---using the Time Series Benchmark Suite (TSBS)\footnote{\url{https://github.com/timescale/tsbs}} developed by the Timescale team;
   \autoref{fig:tsdb} reports the results.

QuestDB sustained the highest ingestion rates, reaching $2.66\times10^{6}$ metrics per second on an 18-core machine, outperforming InfluxDB~v2. Moreover, unlike TimescaleDB, whose throughput plateaus, QuestDB's throughput scales almost linearly with core count over the tested range.
In the infrastructure under study, comprising approximately 70 nodes (\autoref{sec:dataset}), the deployment sustained $\approx 3.1\times10^{5}$ metric values per second. Comparing this load against the benchmarked capacity yields a headroom of at least $8.5\times$, indicating that a sufficiently provisioned instance (i.e., doubling the number of cores) could single-handedly support a deployment an order of magnitude larger, on the order of a thousand nodes.
Each host pushes an independent, host-keyed stream, and no relationship needs to be preserved across hosts; hosts can therefore be freely partitioned across multiple TSDB instances.
Sharding nodes across multiple TSDB instances and merging their dumps in the lake scales capacity linearly with the number of instances, yielding virtually unbounded horizontal scalability.


\section{Dataset Description}
\label{sec:dataset}

The curated dataset is published on Zenodo under the Creative Commons Attribution 4.0 license: each table is an individual Zenodo repository with its metadata and column-level documentation, and an index record links all tables into a unified collection \cite{dataset-zenodo}.

\paragraph{Instrumented infrastructure}
The measurement campaign was conducted at the Orfeo Datacenter, a production research facility located in Trieste, Italy, and operated by AREA Science Park, which serves both a traditional HPC user base and a growing set of cloud-native research services.
The instrumented infrastructure comprises approximately 70 nodes spanning four coexisting subsystems, all of which are covered by the dataset: an HPC partition of 39 nodes organized in seven hardware classes; a 25-node Ceph distributed-storage cluster complemented by a Dell PowerVault SAN; a cloud partition of 10 Kubernetes nodes hosting shared research services, most notably a production LLM inference endpoint~\cite{vllm}; and the network and mission critical IT equipment, which cannot host the collection agent and are therefore instrumented externally at the PDU level (\autoref{req:R4}).
A per-node hardware inventory of all four subsystems is reported in Appendix~\ref{appendix-hardware}.

\paragraph{Campaign}
The dataset covers 28 days of continuous operation, with the facility in normal production use.
All node-level tables collect data at 1~Hz, with the notable exception of the \texttt{pdu\_power} and \texttt{energy\_meter} tables, which are sampled every 30~s\footnote{
    Unlike node-level metrics, which can be collected at high frequency, the latency of SNMP on the PDUs and of the meters' REST interface makes faster polling unreliable.
}.
The campaign produced 2.3~TB of raw data, published as 366~GB of Parquet files totaling more than $2.4\times10^{10}$ rows, plus two context tables.

\subsection{Dataset tables}
\label{subsec:tables}

\autoref{tab:tables} lists them together with their source interface and content; four groups emerge:

\paragraph{Kernel counters} (\texttt{cpu}, \texttt{mem}, \texttt{net}, \texttt{diskio}, \texttt{infiniband}) which describe load: per-core usage, memory occupancy, per-interface traffic, per-device I/O operations, bytes, latencies and merge counts, and InfiniBand port counters. The last three are monotonic counters, so rates are obtained by differencing consecutive samples.

\paragraph{Hardware and sensor readings} (\texttt{turbostat}, \texttt{ipmi\_power}, \texttt{ipmi\_sensor}, \texttt{nvidia\_smi}) describe the physical state: per-core frequencies and package power from model-specific registers (RAPL on Intel, per-core equivalents on AMD Zen), per-PSU and whole-node power from the BMC\footnote{The absolute accuracy of BMC power readings is vendor-dependent; comparisons against external wattmeters report deviations on the order of 5\% \cite{automaticEnergy}. PSU-level readings also enable load-dependent PSU-efficiency modeling \cite{psuEff}.}, component temperatures, fan speeds---a direct proxy for the on-board cooling power, and a useful input for CFD and digital-twin models \cite{cfd}---voltages and currents, and per-GPU power, utilization, memory, clocks, and temperature.

\paragraph{Facility readings} (\texttt{pdu\_power}, \texttt{energy\_meter}) cover per-socket rack power and the electrical-line measurements from which the PUE is derived.

\paragraph{context tables} (\texttt{slurm\_job\_table}, \texttt{llm\_api\_usage}): link telemetry to workloads. 
The first one is a dump of the scheduler database with requested resources, assigned nodes, and execution timestamps for every job,while the last one contains the accounting records of the LLM service with per-request token counts, model, and timestamp.
Joining context tables with telemetry by hostname and time interval is what enables the per-job and per-service attributions. 

\begin{table}[htbp]
\centering
\renewcommand{\arraystretch}{1.15}
\caption{List of tables belonging to the dataset. All telemetry tables are sampled at 1~Hz except \texttt{pdu\_power} and \texttt{energy\_meter} (30~s). Row counts refer to the 28-day campaign.}
\label{tab:tables}
\begin{tabular}{@{}llll@{}}
\toprule
\textbf{Table} & \textbf{Source} & \textbf{Content} & \textbf{Rows} \\
\midrule
cpu & /proc/stat & per-core usage times & 8.72E9 \\
turbostat & MSR/RAPL & per-core freq.\ \& power & 8.49E9 \\
mem & /proc/meminfo & memory occupancy & 1.25E8 \\
net & /proc/net/dev & per-interface traffic & 3.47E8 \\
diskio & /proc/diskstats & per-device I/O activity & 9.52E8 \\
infiniband & sysfs & per-port IB counters & 5.08E8 \\
nvidia\_smi & NVML & per-GPU power \& load & 5.41E7 \\
ipmi\_power & IPMI/BMC & node \& per-PSU power & 8.60E8 \\
ipmi\_sensor & IPMI/BMC & temp., fans, volt., curr. & 4.32E9 \\
pdu\_power & SNMP & per-socket rack power & 1.51E7 \\
energy\_meter & REST & facility electrical lines & 1.24E6 \\
\midrule
slurm\_job\_table & Slurm DB & job resources \& times & per job \\
llm\_api\_usage & vLLM gateway & tokens, model, time & per req. \\
\bottomrule
\end{tabular}
\end{table}

\subsection{Data-quality validation}
\label{subsec:quality}

Since the dataset's purpose is quantitative modeling, we validate the two properties analyses depend on: nothing is missing (\autoref{req:R3}), and timestamps are aligned (\autoref{req:R1}, \autoref{req:R2}).
Both checks group each table's samples into 1-second bins.

\begin{table}[htbp]
    \centering
    \renewcommand{\arraystretch}{1.2}
    \caption{
    Rows per 1-second bin (percentiles, extrema, and standard deviation) and total row count, per telemetry table. A row count that remains constant across bins indicates the absence of missing data.
    }
    \label{tab:bins}
    \begin{tabular}{lccccccc}
\toprule
        \multirow{2}{*}{\begin{tabular}{@{}l@{}}\textbf{Table} \\ \textbf{Name}\end{tabular}} & \multicolumn{6}{c}{\textbf{Rows per 1\,s bin}} & \multirow{2}{*}{\begin{tabular}{@{}c@{}}\textbf{Row} \\ \textbf{Count}\end{tabular}} \\
        \cmidrule(lr){2-7}
        & \textbf{\textit{25th}} & \textbf{\textit{50th}} & \textbf{\textit{75th}} & \textbf{\textit{Min}} & \textbf{\textit{Max}} & \textbf{\textit{Std}} & \\
        \midrule
        ipmi\_sensor & 1811 & 1811 & 1935 & 15 & 2125 & 86.2 & 4.32E9 \\
        diskio & 409 & 409 & 409 & 2 & 541 & 7.1 & 9.52E8 \\
        net & 150 & 150 & 150 & 1 & 165 & 5.7 & 3.47E8 \\
        ipmi\_power & 342 & 384 & 384 & 3 & 398 & 22.6 &  8.60E8 \\
        pdu\_power & 195 & 195 & 195 & 2 & 195 & 1.5 & 1.51E7 \\
        infiniband & 220 & 220 & 220 & 2 & 244 & 9.22 & 5.08E8 \\
        nvidia\_smi & 16 & 24 & 32 & 1 & 40 & 7.8 & 5.41E7 \\
        turbostat &  3580 & 3675 & 3751 & 7 & 4034 & 181.9 & 8.49E9 \\
        cpu &  3781 & 3781 & 3781 & 1 & 3978 & 155.5 & 8.72E9 \\
        mem &  61 & 61 & 61 & 1 & 69 & 1.90 & 1.25E8 \\
        energy\_meter &  16 & 16 & 16 & 16 & 32 & 0.14 & 1.24E6 \\
        \bottomrule
    \end{tabular}
\end{table}

\paragraph{Completeness}
\autoref{tab:bins} reports the distribution of the number of rows falling in each bin.
The count in a bin equals the number of measures concurrently gathered---one row per core for \texttt{cpu} and \texttt{turbostat}, one per block device for \texttt{diskio}, one per PDU socket for \texttt{pdu\_power}, and so on---
so the criterion for completeness is that the count stays constant over the one second large bin.
For most tables the three quartiles coincide exactly and standard deviations are small relative to the medians (5.5\% for the worst case: \texttt{ipmi\_power}): the population of reporting series was stable throughout the campaign.

\begin{table}[htbp]
    \centering
    \renewcommand{\arraystretch}{1.2}
    \caption{Timestamp alignment: distribution of the offset (in milliseconds) between each sample's timestamp and the nearest rounded second, per telemetry table. Small, stable offsets mean every sample is unambiguously assignable to its 1-second bin.}
    \label{tab:align}
    \begin{tabular}{lcccccc}
    \toprule
        \multirow{2}{*}{\textbf{Table Name}} & \multicolumn{6}{c}{\textbf{Offset from bin center (ms)}} \\
        \cmidrule(lr){2-7}
        & \textbf{\textit{25th}} & \textbf{\textit{50th}} & \textbf{\textit{75th}} & \textbf{\textit{Min}} & \textbf{\textit{Max}} & \textbf{\textit{Std}} \\
        \midrule
        ipmi\_sensor & 178 & 236 & 263 & 0 & 500 & 56.57 \\
        diskio & 3 & 5 & 6 & 1 & 48 & 2.17  \\
        net & 1 & 2 & 3 & 0 & 500 & 2.72  \\
        ipmi\_power & 118 & 130 & 180 & 0 & 500 & 48.48  \\
        pdu\_power & 273 & 353 & 423 & 0 & 500 & 106.83  \\
        infiniband & 5 & 8 & 19 & 0 & 500 & 18.93  \\
        nvidia\_smi & 0 & 0 & 0 & 0 & 0 & 0  \\
        turbostat &  124 & 249 & 374 & 0 & 500 & 144.89  \\
        cpu &  1 & 2 & 2 & 0 & 500 & 2.30  \\
        mem &  1 & 1 & 2 & 1 & 500  & 2.96 \\
        energy\_meter &  24 & 25 & 27 & 19 & 408 & 3.849 \\
        \bottomrule
    \end{tabular}
\end{table}

\paragraph{Alignment}
\autoref{tab:align} reports how far each sample's timestamp lies from the nearest second.
Kernel-sourced tables (\texttt{cpu}, \texttt{mem}, \texttt{net}, \texttt{diskio}, \texttt{infiniband}) are stamped within single-digit milliseconds of the ideal grid.
Sources behind slow query protocols---IPMI for \texttt{ipmi\_sensor} and \texttt{ipmi\_power}, SNMP for \texttt{pdu\_power}, model-specific-register sweeps for \texttt{turbostat}---show larger but bounded and stable offsets, inherent to the protocol round-trip rather than to clock error.
This jitter is also what widens their per-bin count distribution in \autoref{tab:bins}: a reading occasionally lands across a bin edge, momentarily moving one row from one bin to the next.
The practical consequence for dataset users is that joins on the 1-second grid are safe for all tables, and a two-second window absorbs the residual edge effects of the slow-protocol tables.


\section{Dataset in action: Case Studies}
\label{sec:cases}

We now demonstrate the analytical capability of the dataset with two studies of workloads that node-local instrumentation fails to fully characterize: distributed I/O, whose energy is consumed on storage servers far from the client application (\autoref{sec:io}), and LLM inference served as a shared multi-tenant service (\autoref{sec:llm}).
As it will be clear in the following lines of this section, the greatest benefit gained using this fine-grained data is the exact magnitudes, component-level breakdowns, and per-workload attributions that aggregate or accelerator-local accounting cannot provide.

\subsection{Energy efficacy of distributed I/O}
\label{sec:io}
\subsubsection{Methodology and metrics}
\label{sec:io-method}

We characterize the power consumption of I/O on the production Ceph file system by combining data regarding the storage activity (\texttt{diskio}), their electrical power (\texttt{ipmi\_power}, \texttt{ipmi\_sensor}, \texttt{turbostat}) and the application-level bandwidth measured on both compute and storage nodes (\texttt{net}).
The synthetic workload is performed through the IOR benchmark, executed on two compute nodes against two CephFS data pools: mechanical drives (HDD) with 6+2 erasure coding, and NVMe drives with $2\times$ replication.

We tested sequential and random patterns, 4~KB and 4~MB transfer sizes, and the POSIX and MPI-IO backends, repeating each configuration 10 times; benchmark windows are delimited using the Slurm job records.
The runs executed on the production system, isolated up to the background noise of uninterruptible services.

We define two \emph{efficacies}, both normalized by the power $p$ of the storage servers during the run, and expressed in MB/s per W\footnote{Dimensionally, MB/s per W is equivalent to MB/J, since $1,\mathrm{W} = 1,\mathrm{J/s}$. Thus, both formulations quantify the amount of data transferred per unit of energy.}.

\begin{equation}
\eta_{\mathrm{gross}} = \frac{B_{\mathrm{dev}}}{p},
\qquad
\eta_{\mathrm{net}} = \frac{B_{\mathrm{app}}}{p},
\label{eq:efficacy}
\end{equation}
where $B_{\mathrm{dev}}$ is the aggregate bandwidth actually served by the storage devices (from \texttt{diskio} counters) and $B_{\mathrm{app}}$ is the bandwidth observed by the application (from IOR).
The gross efficacy measures how much device-level bandwidth is delivered for each watt of power drawn by the storage servers \cite{exploringIO}; the net efficacy measures how much \emph{useful} application bandwidth is delivered for each watt drawn.
Their ratio is a genuine, dimensionless \emph{efficacies ratio},

\begin{equation}
\varepsilon = \frac{\eta_{\mathrm{net}}}{\eta_{\mathrm{gross}}} = \frac{B_{\mathrm{app}}}{B_{\mathrm{dev}}},
\label{eq:conversion}
\end{equation}
in which power cancels: $\varepsilon$ is the fraction of device traffic that serves application data, while $1/\varepsilon$ can be seen as the I/O amplification factor of the file system for that workload.

Analogous quantities are defined for operation rates (IOPS per Watt); unlike its bandwidth counterpart, the operation-count ratio can exceed one, as the file system may aggregate many small application operations into fewer, larger device operations.

\subsubsection{Cluster-level measurement analysis}

\begin{figure}[htbp]
    \centering
    \includegraphics[width=\columnwidth]{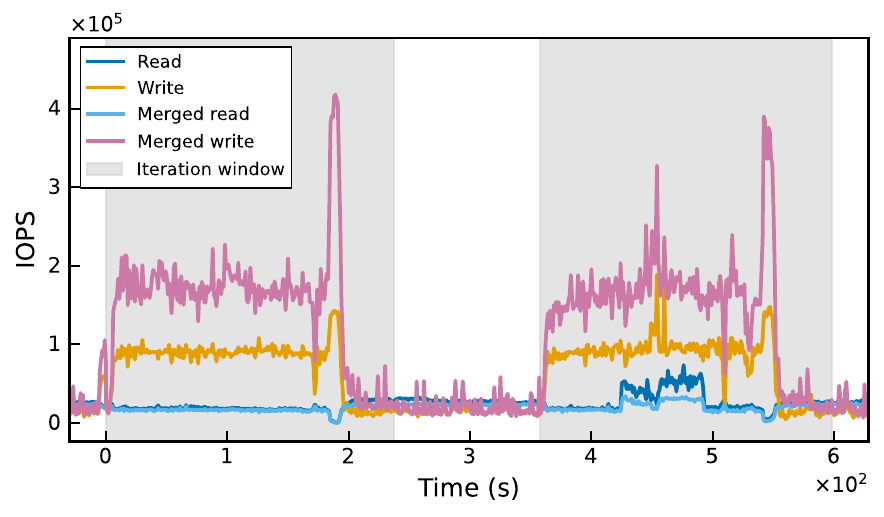}
    \caption{Aggregate device operations (read, write, and merged operations per second) across the 23 Ceph storage servers (389 data drives) during two iterations of the IOR sequential-write MPI-IO 4~MB benchmark. Vertical markers delimit the iterations; note the flush bursts at the end of each write phase.}
    \label{fig:ior-diskio}
\end{figure}

\begin{figure}[htbp]
    \centering
    \includegraphics[width=\columnwidth]{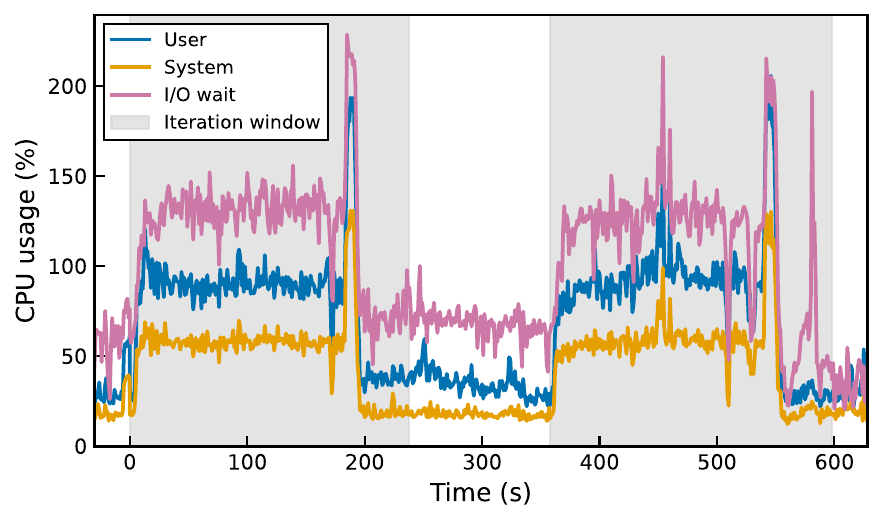}
    \caption{Aggregate CPU usage (user, system, and iowait) across the 23 Ceph storage servers during the same two IOR iterations of \autoref{fig:ior-diskio}. The servers are I/O-bound: iowait dominates throughout the runs.}
    \label{fig:ior-cpu}
\end{figure}

\begin{figure}[htbp]
    \centering
    \includegraphics[width=\columnwidth]{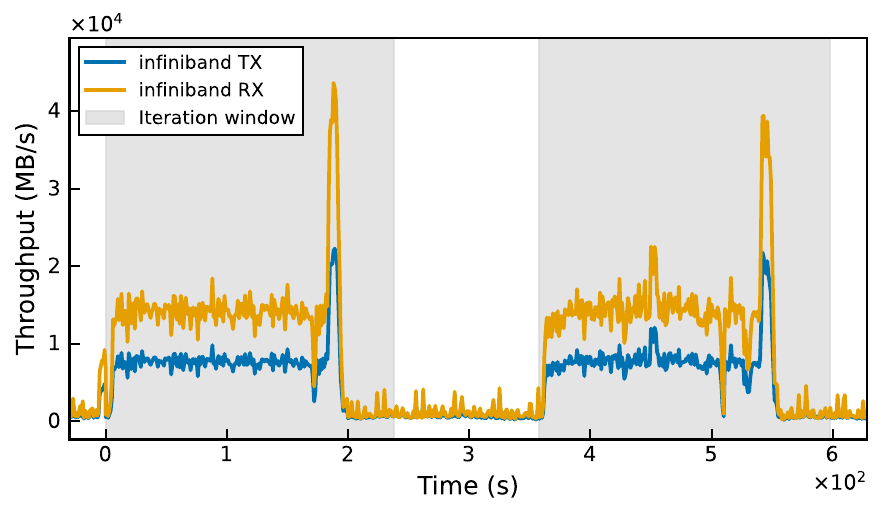}
    \caption{Aggregate network throughput of the 23 Ceph storage servers during the same two IOR iterations. Traffic rides entirely on InfiniBand, with receive roughly twice transmit, consistent with the redistribution of 2 replicas across servers.}
    \label{fig:ior-net}
\end{figure}

\begin{figure}[htbp]
    \centering
    \includegraphics[width=\columnwidth]{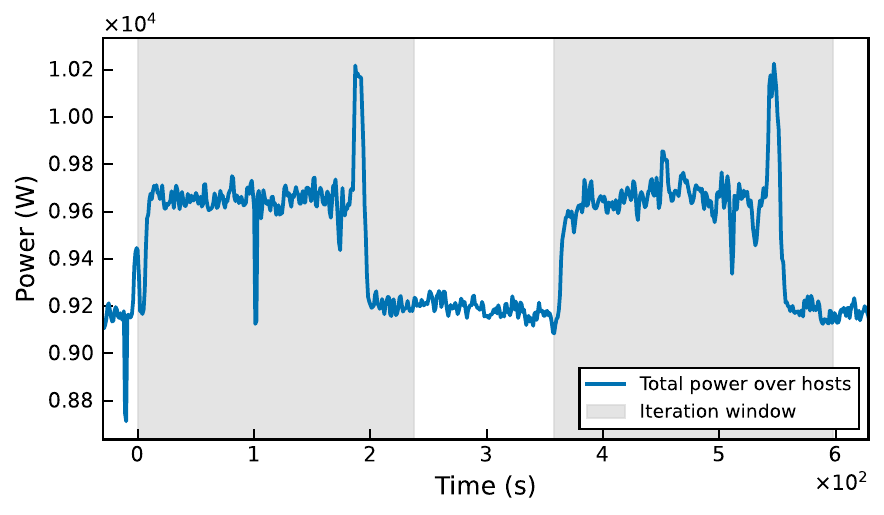}
    \caption{Aggregate electrical power of the 23 Ceph storage servers during the same two IOR iterations. The benchmark raises consumption by only $\approx 5\%$ over the $\approx 9.2$~kW baseline: the energy cost of I/O is dominated by the always-on draw of the storage subsystem.}
    \label{fig:ior-power}
\end{figure}

Figures \ref{fig:ior-diskio}--\ref{fig:ior-power} show the storage-cluster telemetry during two consecutive iterations of the sequential-write MPI-IO 4~MB configuration, and illustrate the character of the raw data.
\autoref{fig:ior-diskio} shows device operations: write IOPS plateau near $10^{5}$ with merged writes roughly twice as high, and each iteration ends with a sharp burst---consistent with the flush of outstanding buffers at file close---reaching $4\times10^{5}$ IOPS.
\autoref{fig:ior-cpu} shows that the storage servers are I/O-bound rather than compute-bound during the runs: iowait rises far above user and system time.
\autoref{fig:ior-net} shows the same experiment from the network point of view: all data flows on the InfiniBand fabric ($\approx 15$~GB/s aggregate receive at plateau), with receive traffic roughly double transmit traffic, as expected on a replica 2 pool where each object arrives once from the client and is then redistributed as coded chunks among peer servers.
\autoref{fig:ior-power} shows the electrical consequence: the 23 storage servers idle at $\approx 9.2$~kW and rise only to $\approx 9.65$~kW under load, with transient peaks above 10~kW.
This last observation shapes the whole analysis: the storage subsystem's baseline draw dominates its marginal draw ($\approx 5\%$), so the energy cost of I/O is primarily the cost of \emph{owning} the subsystem for the duration of the workload, not of the additional watts it induces.

\begin{table*}[t]
	\centering
	\renewcommand{\arraystretch}{1.2}
	\caption{Energy efficacy of the Ceph file system on the \textbf{HDD} backend (mechanical-drive pool with 6+2 erasure coding), per IOR scenario and I/O interface (10 repetitions each). Gross efficacy $\eta_{\mathrm{gross}}$ uses the device-level activity of the storage servers (\texttt{diskio}); net efficacy $\eta_{\mathrm{net}}$ uses the application-level bandwidth (IOR); both are normalized by the power of the storage servers. $\varepsilon$ is the net-to-gross conversion (\autoref{eq:conversion}); the operation-count conversion may exceed 1 due to client-side aggregation.}
	\label{tab:io-hdd}
	\begin{tabular}{llrrrrrrr}
		\toprule
		\multirow{2}{*}{\textbf{Scenario}} & \multirow{2}{*}{\textbf{Interface}} & \multicolumn{2}{c}{\textbf{Efficacy (MB/s per W)}} & \multicolumn{2}{c}{\textbf{Efficacy (IOPS per W)}} & \multicolumn{2}{c}{\textbf{Efficacy ratio $\varepsilon$}} & \multirow{2}{*}{\begin{tabular}{@{}c@{}}\textbf{Mean} \\ \textbf{Power (W)}\end{tabular}} \\
		\cmidrule(lr){3-4} \cmidrule(lr){5-6} \cmidrule(lr){7-8}
		& & \textbf{Gross} & \textbf{Net} & \textbf{Gross} & \textbf{Net} & \textbf{BW} & \textbf{Ops} & \\
		\midrule
		Seq. write 4KB  & POSIX & 0.660 & 0.001 & 42.644 & 0.215 & 0.001 & 0.005 & 9454 \\
		Seq. read 4KB   & POSIX & 0.270 & 0.035 & 14.630 & 9.007 & 0.131 & 0.616 & 9793 \\
		Seq. write 4MB  & POSIX & 0.853 & 0.413 & 25.547 & 0.103 & 0.484 & 0.004 & 9463 \\
		Seq. read 4MB   & POSIX & 0.826 & 0.603 & 17.852 & 0.151 & 0.730 & 0.008 & 9471 \\
		Rand. write 4KB & POSIX & 0.788 & 0.001 & 25.295 & 0.129 & 0.001 & 0.005 & 9465 \\
		Rand. read 4KB  & POSIX & 0.247 & 0.001 &  8.313 & 0.247 & 0.004 & 0.030 & 9278 \\
		\midrule
		Seq. write 4KB  & MPIIO & 0.854 & 0.369 & 24.636 &  94.480 & 0.429 &  3.814 & 9452 \\
		Seq. read 4KB   & MPIIO & 1.102 & 0.823 & 20.520 & 210.762 & 0.747 & 10.265 & 9517 \\
		Seq. write 4MB  & MPIIO & 1.162 & 0.650 & 27.632 &   0.163 & 0.556 &  0.006 & 9585 \\
		Seq. read 4MB   & MPIIO & 1.055 & 0.858 & 19.526 &   0.214 & 0.813 &  0.011 & 9513 \\
		Rand. read 4KB  & MPIIO & 0.691 & 0.001 & 13.330 &   0.139 & 0.001 &  0.010 & 9421 \\
		\bottomrule
	\end{tabular}
\end{table*}

\begin{table*}[t]
	\centering
	\renewcommand{\arraystretch}{1.2}
	\caption{Energy efficacy of the Ceph file system on the \textbf{NVMe} backend (NVMe pool with $2\times$ replication), per IOR scenario and I/O interface (10 repetitions each). Metrics are defined as in \autoref{tab:io-hdd}: gross efficacy $\eta_{\mathrm{gross}}$ uses the device-level activity of the storage servers (\texttt{diskio}), net efficacy $\eta_{\mathrm{net}}$ uses the application-level bandwidth (IOR), and both are normalized by the power of the storage servers. $\varepsilon$ is the net-to-gross conversion (\autoref{eq:conversion}); the operation-count conversion may exceed 1 due to client-side aggregation.}
	\label{tab:io-nvme}
	\begin{tabular}{llrrrrrrr}
		\toprule
		\multirow{2}{*}{\textbf{Scenario}} & \multirow{2}{*}{\textbf{Interface}} & \multicolumn{2}{c}{\textbf{Efficacy (MB/s per W)}} & \multicolumn{2}{c}{\textbf{Efficacy (IOPS per W)}} & \multicolumn{2}{c}{\textbf{Efficacy ratio $\varepsilon$}} & \multirow{2}{*}{\begin{tabular}{@{}c@{}}\textbf{Mean} \\ \textbf{Power (W)}\end{tabular}} \\
		\cmidrule(lr){3-4} \cmidrule(lr){5-6} \cmidrule(lr){7-8}
		& & \textbf{Gross} & \textbf{Net} & \textbf{Gross} & \textbf{Net} & \textbf{BW} & \textbf{Ops} & \\
		\midrule
		Seq. write 4KB  & POSIX & 1.020 & 0.109 & 167.620 & 27.936 & 0.107 & 0.166 & 9831 \\
		Seq. read 4KB   & POSIX & 0.560 & 0.196 &  59.876 & 50.268 & 0.351 & 0.839 & 9328 \\
		Seq. write 4MB  & POSIX & 3.152 & 1.495 &  57.388 &  0.374 & 0.473 & 0.006 & 9586 \\
		Seq. read 4MB   & POSIX & 1.572 & 1.308 & 291.779 &  0.327 & 0.832 & 0.001 & 9306 \\
		Rand. write 4KB & POSIX & 0.864 & 0.093 & 160.258 & 23.873 & 0.108 & 0.149 & 9372 \\
		Rand. read 4KB  & POSIX & 0.460 & 0.180 &  53.493 & 46.038 & 0.391 & 0.860 & 9294 \\
		\midrule
		Seq. write 4KB  & MPIIO & 0.985 & 0.375 &  22.616 &  96.068 & 0.380 & 4.233 & 9217 \\
		Seq. read 4KB   & MPIIO & 1.532 & 1.235 & 275.360 & 316.041 & 0.805 & 1.146 & 9271 \\
		Seq. write 4MB  & MPIIO & 1.831 & 0.862 &  34.362 &   0.215 & 0.468 & 0.006 & 9391 \\
		Seq. read 4MB   & MPIIO & 1.537 & 1.177 & 264.772 &   0.294 & 0.765 & 0.001 & 9287 \\
		Rand. read 4KB  & MPIIO & 1.435 & 0.002 & 253.408 &   0.413 & 0.001 & 0.002 & 9257 \\
		\bottomrule
	\end{tabular}
\end{table*}

\subsubsection{Device-level view: gross efficacy}

 \autoref{tab:io-hdd} and \autoref{tab:io-nvme} report the results for the HDD and NVMe pools respectively.
At the device level, NVMe is more energy-efficacious than HDD in nearly every scenario: gross efficacy ranges from 0.247 to 1.162~MB/s per W on mechanical drives and from 0.460 to 3.152~MB/s per W on NVMe.
The gap is widest where mechanical media are weakest: for sequential 4~MB writes with POSIX, NVMe delivers $3.69\times$ the efficacy of HDD (3.152 vs.\ 0.853~MB/s per W), and random 4~KB reads collapse on HDD (0.247~MB/s per W) because seek latency leaves the drives---and the servers powering them---mostly waiting.
For some aggregation-friendly patterns, however, the two pools are nearly equivalent: random 4~KB writes with POSIX achieve 0.788~MB/s per W on HDD against 0.864 on NVMe, and sequential 4~KB writes with MPI-IO reach 0.854 against 0.985~MB/s per W, since batched, deep-queue writes let mechanical drives stream near their sequential rate.
We can conclude that even if NVMe can reach higher efficacy, HDD can still be competitive for some I/O pattern.

\subsubsection{Application-level view: net efficacy}

A markedly different picture emerges when considering net efficacy, which accounts for the bandwidth actually delivered to the \emph{application}, rather than the bandwidth generated at the device level, per watt consumed.

Best cases are large sequential transfers: up to 0.858~MB/s per W on HDD (sequential 4~MB reads, MPI-IO) and 1.495~MB/s per W on NVMe (sequential 4~MB writes, POSIX).
Worst cases are dramatic: every 4~KB POSIX write pattern on HDD, and random 4~KB reads on HDD, deliver $\approx 0.001$~MB/s per W---three orders of magnitude below the gross efficacy of the same runs.
The backend matters as much as the medium: with MPI-IO, sequential 4~KB writes on HDD jump from 0.001 to 0.369~MB/s per W, and on NVMe random 4~KB reads reach 0.180~MB/s per W with POSIX but only 0.002 with MPI-IO.
Net efficacy is the quantity that should enter any user-facing energy accounting, since it reflects delivered service; comparing it with gross efficacy is what exposes where the missing energy goes.

\subsubsection{Efficacies ratio analysis}

The efficacies ratio $\varepsilon$ of \autoref{eq:conversion} localizes the losses.
Small POSIX writes on the erasure-coded HDD pool show $\varepsilon = 0.001$: the file system moves a thousand device bytes per useful byte.
This is write amplification at its worst---sub-stripe 4~KB writes force read-modify-write cycles on full erasure-coded stripes, on top of journaling and metadata traffic---and it explains the collapse of net efficacy observed above.
The MPI-IO backend repairs precisely this pathology: collective buffering aggregates client requests into large, stripe-aligned writes, lifting $\varepsilon$ to 0.429 for the same nominal workload.
The same aggregation is visible in the operation-count ratios, which exceed unity (up to 10.265 for sequential 4~KB reads on HDD): many small application operations complete against few large device operations.
Conversely, random 4~KB reads with MPI-IO are inefficient on both pools ($\varepsilon = 0.001$): random access defeats collective buffering while still paying its coordination cost, and on HDD every miss additionally pays read amplification from read-ahead and allocation granularity ($\varepsilon = 0.004$ with plain POSIX, against 0.391 on NVMe).

Large sequential transfers approach the redundancy-imposed ceiling: $\varepsilon$ reaches 0.813 (HDD, MPI-IO reads) and 0.832 (NVMe, POSIX reads).
For writes, redundancy schemes set the achievable bound, and here the dataset yields a result worth stressing: on sequential writes with MPI-IO, the mechanical pool with erasure coding outperforms the NVMe pool with replication ($\varepsilon =$ 0.429 and 0.556 for 4~KB and 4~MB, against 0.380 and 0.468)---because 6+2 erasure coding writes $\approx 1.33$ redundant bytes per user byte while $2\times$ replication writes two.

\subsection{Component-resolved energy accounting of LLM inference}
\label{sec:llm}
\begin{table*}[htbp]
    \centering
    \renewcommand{\arraystretch}{1.2}
    \caption{Performance and per-token energy of three LLMs served by vLLM on an 8$\times$H100 node, for varying GPU allocations. Energy is reported in Wh per 1{,}000 generated tokens, decomposed as in \autoref{eq:llm}: active and idle GPUs (NVML), CPU (RAPL), Other (residual: fans, NVSwitch, NVMe, motherboard, RAM), and the whole-node System total (IPMI). The Facility column applies the PUE measured contemporaneously with each run (1.35--1.41).}
    \label{tab:llm}
    \begin{tabular}{lrrrrrrrrrr}
        \toprule
        \multirow{2}{*}{\textbf{Model}} & \multicolumn{2}{c}{\textbf{GPUs}} & \multirow{2}{*}{\textbf{Tokens/s}} & \multirow{2}{*}{\begin{tabular}[c]{@{}c@{}}\textbf{GPU}\\ \textbf{Usage (\%)}\end{tabular}} & \multicolumn{5}{c}{\textbf{Energy / 1k Tokens (Wh)}} & \multirow{2}{*}{\begin{tabular}[c]{@{}c@{}}\textbf{Facility}\\ \textbf{PUE-corr.\ (Wh)}\end{tabular}} \\
        \cmidrule(lr){2-3} \cmidrule(lr){6-10}
        & \textbf{Act.} & \textbf{Idle} & & & \textbf{Busy\ GPU} & \textbf{Idle GPU} & \textbf{CPU} & \textbf{Other} & \textbf{System} & \\
        \midrule
        \multirow{3}{*}{GPT-OSS-120B}
         & 2 & 6 & 1478.31 & 100.00 & 0.18 & 0.08 & 0.09 & 0.23 & 0.58 & 0.81 \\
         & 4 & 4 & 1501.70 &  97.99 & 0.26 & 0.05 & 0.09 & 0.23 & 0.63 & 0.89 \\
         & 8 & 0 & 1491.43 &  94.74 & 0.42 & 0.00 & 0.09 & 0.26 & 0.77 & 1.07 \\
        \midrule
        \multirow{3}{*}{Llama-4-Scout-17B}
         & 2 & 6 &  928.01 &  93.93 & 0.26 & 0.13 & 0.14 & 0.35 & 0.88 & 1.20 \\
         & 4 & 4 & 1048.44 &  89.81 & 0.36 & 0.08 & 0.13 & 0.32 & 0.89 & 1.21 \\
         & 8 & 0 &  973.74 &  74.18 & 0.63 & 0.00 & 0.14 & 0.40 & 1.17 & 1.60 \\
        \midrule
        \multirow{2}{*}{Qwen3-VL-235B-Thinking}
         & 4 & 4 &  799.73 &  94.57 & 0.56 & 0.10 & 0.16 & 0.45 & 1.27 & 1.72 \\
         & 8 & 0 &  964.87 &  88.01 & 0.75 & 0.00 & 0.14 & 0.40 & 1.29 & 1.77 \\
        \bottomrule
    \end{tabular}
\end{table*}

\subsubsection{Motivation and methodology}
\label{sec:llm-method}

Modern AI research pipelines, especially those involving autonomous agents, explainability studies, LLM benchmarking, embedding generation, and synthetic data creation, rely heavily on model inference.
Historically, individual models were deployed directly within the traditional HPC queue, resulting in significant resource inefficiencies and redundant model deployments.
However, it is increasingly common  migrating from per-user deployments in the HPC queue to a centralized, multi-tenant service: models served on Kubernetes behind an OpenAI-compatible API.
Centralization improves utilization but breaks per-node energy attribution: multiple users share models on shared hardware, so the natural accounting unit becomes the \emph{token}.

We benchmarked three production models (GPT-OSS-120B, Llama-4-Scout-17B, and Qwen3-VL-235B-Thinking) served by vLLM on a Kubernetes node with eight NVIDIA H100 GPUs and two Xeon Platinum 8480+ CPUs\footnote{The node appears in the dataset under the hostname \texttt{dgx005}.}, scaling each model across 2, 4, or 8 GPUs.
Load was generated with Locust emulating 20 concurrent users, each drawing from 200 fixed prompts with randomized maximum answer lengths between 1000 and 2000 tokens, prompting continuously for 15 minutes per configuration.

Energy is reported in Wh per 1000 generated tokens and decomposed by component:
\begin{equation}
E_{\mathrm{sys}} = E_{\mathrm{CPU}} + E_{\mathrm{GPU}}^{\mathrm{busy}} + E_{\mathrm{GPU}}^{\mathrm{idle}} + E_{\mathrm{other}},
\label{eq:llm}
\end{equation}

\begin{equation}
E_{\mathrm{fac}} = \mathrm{PUE}(t)\, E_{\mathrm{sys}},
\label{eq:llm-pue}
\end{equation}

where $E_{\mathrm{sys}}$ is the whole-node energy from IPMI, CPU energy comes from RAPL (\texttt{turbostat}), and per-GPU energies from NVML (\texttt{nvidia\_smi}), split between the GPUs assigned to the model and the idle ones.
$E_{\mathrm{other}}$ is the residual: components that cannot be measured individually---node cooling fans (substantial in an 11~kW-class machine), the NVSwitch fabric, four on-board NVMe drives, the motherboard, and 1~TB of RAM.

\begin{figure}[htbp]
    \centering
    \includegraphics[width=\columnwidth]{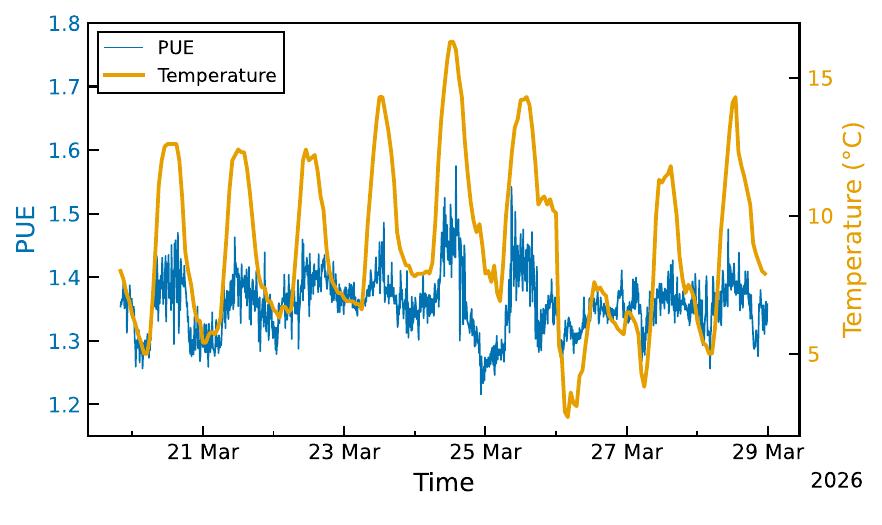}
    \caption{Datacenter Power Usage Effectiveness (PUE) 
    during 10 days of observation.
    The external air temperature from the nearest weather station (45.64$^\circ$\,N, 13.85$^\circ$\,E), queried via the open-meteo API \cite{meteo}, is also displayed, illustrating a clear relationship between PUE and external temperature.}
    \label{fig:pue-trend}
\end{figure}

The facility-level cost $E_{\mathrm{fac}}$ deserves justification, since it is the step most often omitted.
Cooling and power-delivery overhead constitute genuine energy expenditure incurred on behalf of the workload; consequently, any carbon or cost accounting that excludes this term underestimates the true footprint.
We therefore adopt the PUE measured directly by our own pipeline, contemporaneously with each benchmark run, rather than a nameplate or annualized value, since PUE is far from constant: over the course of the measurement campaign, the facility overhead ranged from 25\% to 60\% (i.e., $\mathrm{PUE} = 1.25$--$1.60$) and is influenced by several external factors, such as ambient temperature, as shown in \autoref{fig:pue-trend}.

This implies that the facility-side cost of an identical workload may vary by as much as a quarter, depending solely on the time at which it is executed. During the benchmark windows, the measured PUE ranged between 1.35 and 1.41.
For that reason, we propose a time-resolved correction using the contemporaneously measured value as a more accurate approach for converting IT-side measurements into facility-side accounting.

\subsubsection{Results}
\label{sec:llm-results}

\autoref{tab:llm} reports throughput, GPU utilization, and the energy breakdown of \autoref{eq:llm} and the computed value of \autoref{eq:llm-pue}.
Two observations stand out.

\paragraph{GPUs are less than half of the true cost}
Active GPUs never dominate the facility-level energy: their share of $E_{\mathrm{sys}}$ peaks at 58\% (Qwen3-VL on 8 GPUs) and falls to 31\% for GPT-OSS-120B on 2 GPUs; against $E_{\mathrm{fac}}$ the shares drop further.
The residual $E_{\mathrm{other}}$ alone accounts for 34--40\% of system energy across configurations, and CPU energy is far from negligible: up to half the active-GPU energy for GPT-OSS-120B on 2 GPUs (0.09 vs.\ 0.18~Wh per 1{,}000 tokens).
Consequently, an accounting based on GPU counters alone underestimates the facility-level cost by a factor of $2.4\times$ to $4.6\times$ across our configurations---e.g., 0.18 vs.\ 0.81~Wh per 1{,}000 tokens for GPT-OSS-120B on 2 GPUs, and 0.75 vs.\ 1.77 for Qwen3-VL-235B on 8 GPUs.

\paragraph{Energy scalability}
Scaling GPT-OSS-120B from 2 to 8 GPUs leaves throughput essentially unchanged ($\approx 1{,}480$--$1{,}500$ tokens/s) while multiplying active-GPU energy by 2.33; for this model, the additional GPUs buy nothing but heat, and the most efficient deployment is the smallest one, with the unused GPUs freed for other services.
The large Qwen3-VL-235B model behaves differently: going from 4 to 8 GPUs improves throughput by 20\% and leaves the per-token facility cost nearly flat (1.72 vs.\ 1.77~Wh).
Right-sizing therefore requires per-model measurement---and the breakdown shows why aggregate metrics mislead: as GPU count grows, utilization percentages fall (94.74\% to 74.18\% for Llama-4-Scout from 2 to 8 GPUs) while absolute GPU energy rises, and utilization does not rank models by power either (Llama-4-Scout runs at lower utilization than GPT-OSS-120B yet consistently draws more energy per token).


\section{Conclusions}
\label{sec:conclusion}

We presented two complementary artifacts, both public.
The first is a measurement pipeline built entirely from off-the-shelf open-source components synchronized across the infrastructure with lossless buffered collection, and coverage extending from kernel counters to facility power meters with an invisible collection overhead (approximately it can be estimated in our case around $\approx$13.58 KB/s\footnote{
  2.3 TB over 28 days, $\approx$ 70 instrumented nodes, i.e., 0.95 MB/s aggregated.
  }).
The second artifact is a 28-day, $2.4\times10^{10}$-row dataset from a production research facility---compute, storage, network, cloud services, and facility metering in one consistent, validated schema, joined by scheduler and LLM-billing context tables.

The case studies demonstrate what this combination enables.
On distributed I/O, simultaneous client-side and device-side measurement separated gross from net energy efficacy, quantified I/O amplification and showed that the redundancy policy, more than the storage medium, governs write-path efficiency.
On LLM inference, whole-node decomposition with time-resolved PUE correction showed that GPU-only accounting understates facility cost by $2.4\times$--$4.6\times$.

Because the pipeline is reproducible from public components and configurations, any HPC facility can deploy it and publish comparable datasets; we hope this lowers the barrier to the cross-site, infrastructure-level energy studies the field currently lacks. 
Future work will focus on leveraging this dataset to construct a dedicated energy model that accurately characterizes I/O operations on distributed storage from computational nodes. By integrating metrics from storage nodes, network traffic, and I/O times on compute nodes, the goal is to establish an energy model that accounts for energy consumption not captured by single-node instrumentation.

\section*{Acknowledgment}

The authors acknowledge the Area Science Park supercomputing platform ORFEO made available for conducting the research reported in this paper and the technical support of the Laboratory of Data Engineering staff. 

I.P. was supported by the Programma Nazionale della Ricerca (PNR) grant J95F21002830001 with the title “FAIR-by-design”. 

As non-native English speakers, the authors used language tools for proofreading, rephrasing, and grammar checking like Grammarly and open-source Large language models (Qwen3.6).
 No AI-model was used to produce the original content in this work.

\bibliographystyle{IEEEtran}
\bibliography{references}

\appendices

\section{Deployment Environment}
\label{appendix-hardware}

In this appendix we present a detailed description of all the hardware partition involved in the measurement campaign described in \autoref{sec:dataset}.

\subsection{HPC Partition}

The HPC partition consists of 39 nodes grouped into seven hardware classes, whose configurations are detailed in \autoref{tab:HPC-pariitions-details}.
Every node is a dual-socket server attached to a low-latency Mellanox InfiniBand fabric.
The classes span several hardware generations from Intel Xeon Gold platforms to recent AMD EPYC (Genoa and Rome) and Intel Xeon Platinum systems.
The GPU-accelerated classes (V100, A100, and H100) additionally host multiple high-end NVIDIA accelerators per node, and provide the compute substrate for the GPU workloads analyzed in \autoref{sec:cases}.

\begin{table}[htbp]
\centering
\caption{Summary of resource across HPC nodes.}
\label{tab:HPC-pariitions-details}
\begin{tabular}{lr}
\toprule
\textbf{Component} & 13 $\times$ \textbf{GENOA nodes} \\
\midrule
Model &  Dell PowerEdge R6625  \\
CPU &  2 $\times$ AMD EPYC 9374F 32C \\
Memory & 16 $\times$ 32 GB DDR5 @ 4800MT/s \\
Infiniband & Mellanox ConnectX-7 @ 200 Gbit/s \\
\toprule
\textbf{Component} & 8 $\times$ \textbf{ROME nodes} \\
\midrule
Model &  Dell PowerEdge R6525  \\
CPU &  2 $\times$ AMD EPYC 7H12 64C \\
Memory & 16 $\times$ 32 GB DDR4 @ 3200 MT/s \\
Infiniband & Mellanox ConnectX-7 @ 100 Gbit/s \\
\toprule
\textbf{Component} & 8 $\times$ \textbf{THIN nodes} \\
\midrule
Model &  Dell PowerEdge R640  \\
CPU &  2 $\times$ Intel Xeon Gold 6126 12C \\
Memory & 12 $\times$ 64 GB DDR4 @ 2666 MT/s \\
Infiniband & Mellanox ConnectX-6 @ 100 Gbit/s \\
\toprule
\textbf{Component} & 2 $\times$ \textbf{FAT nodes} \\
\midrule
Model &  Dell PowerEdge R640  \\
CPU &  2 $\times$ Intel Xeon Gold 6154 18C \\
Memory & 24 $\times$ 64 GB DDR4 @ 2666 MT/s \\
Infiniband & Mellanox ConnectX-6 @ 100 Gbit/s \\
\toprule
\textbf{Component} & 3 $\times$ \textbf{V100 nodes} \\
\midrule
Model &  Dell PowerEdge R740xd  \\
CPU &  2 $\times$ Intel Xeon Gold  6226  12C \\
Memory & 8 $\times$ 32 GB DDR4 @ 2933 MT/s \\
Infiniband & Mellanox ConnectX-6 @ 100 Gbit/s \\
GPU & 2 $\times$ NVIDIA Tesla V100 PCIe 32GB \\
\toprule
\textbf{Component} & 2 $\times$ \textbf{A100 nodes} \\
\midrule
Model &  NVIDIA DGX A100 \\
CPU &  2 $\times$  AMD EPYC 7742 64C \\
Memory & 16 $\times$ 64 GB DDR4 @ 3200 MT/s \\
Infiniband & 8 $\times$ Mellanox ConnectX-6 @ 100 Gbit/s \\
GPU & 8 $\times$ NVIDIA Ampere A100 PCIe 40GB \\
\toprule
\textbf{Component} & 3 $\times$ \textbf{H100 nodes} \\
\midrule
Model &  Dell PowerEdge XE9680 \\
CPU &  2 $\times$  Intel Xeon Platinum 8480+ 56C \\
Memory & 16 $\times$ 64 GB DDR5 @ 4800 MT/s \\
Infiniband & 8 $\times$ Mellanox ConnectX-7 @ 200 Gbit/s \\
GPU & 8 $\times$ NVIDIA Hopper H100 PCIe 80GB \\
\bottomrule
\end{tabular}
\end{table}

\subsection{Storage Infrastructure}

The storage infrastructure is built around a 25-node Ceph cluster comprising 23 object-storage (OSD) servers and 2 metadata servers (MDS), all based on Intel Xeon processors and populated with a heterogeneous mix of mechanical (HDD), SSD, and NVMe drives.
The cluster provides approximately 5~PB of raw capacity, exposed to clients through the CephFS, RBD, and S3 interfaces; its per-node hardware is detailed in \autoref{tab:Storage-partition-details}.
In addition, a Dell PowerVault storage-area network populated with 252 mechanical drives provides roughly 2~PB of long-term capacity exported over NFS.
The Ceph cluster is the subject of the distributed-I/O case study of \autoref{sec:io}.

\begin{table}[htbp]
\centering
\caption{Summary of resource across Ceph cluster nodes.}
\label{tab:Storage-partition-details}
\begin{tabular}{lr}
\toprule
\textbf{Component} & 4 $\times$ \\
\midrule
Model &  Dell PowerEdge R740xd  \\
CPU &  2 $\times$ Intel Gold 5115 10C \\
Memory & 12 $\times$ 16 GB DDR4 @ 2666MT/s \\
SSD & 14 $\times$ Dell DC29P 3.5TB  \\
    & 6 $\times$ Dell FYP5F - 3.5TB \\
NVMe & 2 $\times$ Dell Express Flash PM1725b 1.6TB \\
\toprule
\textbf{Component} & 4 $\times$  \\
\midrule
Model &  Dell PowerEdge R740xd  \\
CPU &  2 $\times$ Intel Gold 5115 10C \\
Memory & 6 $\times$ 16 GB DDR4 @ 2666MT/s \\
HDD & 17 $\times$ HGST HUH721212AL5200 10.7TB  \\
    & 3 $\times$ Thoshiba MG07SCA12TEY 10.7 TB  \\
\toprule
\textbf{Component} & 2 $\times$ \\
\midrule
Model &  Dell PowerEdge R740xd  \\
CPU &  2 $\times$ Intel Gold 5115 10C \\
Memory & 6 $\times$ 16 GB DDR4 @ 2666MT/s \\
HDD & 12 $\times$ Thoshiba MG07SCA12TEY 10.7TB  \\
    & 3 $\times$ Seagate ST12000NM0158 10.7 TB  \\
NVMe & 1 $\times$ Dell Ent NVMe AGN MU AIC 1.6TB \\
\toprule
\textbf{Component} & 3 $\times$ \\
\midrule
Model &  Dell PowerEdge R740xd2  \\
CPU &  2 $\times$ Intel Gold 5215 10C \\
Memory & 6 $\times$ 16 GB DDR4 @ 3200/s \\
HDD & 18 $\times$ Seagate ST16000NM006J 14.6TB  \\
NVMe &  1 $\times$ Dell Ent NVMe AGN MU AIC 1.6TB \\
\toprule
\textbf{Component} & 10 $\times$ \\
\midrule
Model &  Dell PowerEdge R760 - Storage - rev1.0  \\
CPU &  2 $\times$ Intel Gold 6426Y 16C \\
Memory & 12 $\times$ 16 GB DDR5 @ 4800/s \\
HDD & 18 $\times$ WDC WUH722222AL5200  20TB  \\
NVMe &  2 $\times$ Dell NVMe ISE PS1010 RI U.2 16TB \\
\toprule
\textbf{Component} & 2 $\times$ metadataserver \\
\midrule
Model &  Dell PowerEdge R640  \\
CPU &  2 $\times$ Intel Gold 6126 12C \\
Memory & 12 $\times$ 64 GB DDR4 @ 2666/s \\
\bottomrule
\end{tabular}
\end{table}

\subsection{Cloud Partition}

The cloud partition consists of 10 Intel Xeon-based nodes managed by Kubernetes and dedicated to shared research services.
The most resource-intensive of these is a production LLM inference endpoint~\cite{vllm}, served on two nodes whose hardware configuration is identical to that of the H100 and V100 classes reported in \autoref{tab:HPC-pariitions-details}.
This partition is the subject of the inference case study of \autoref{sec:llm}.

\subsection{Network and Perimeter Equipment}

Inter-node communication and external connectivity are provided by a dedicated set of switches and gateways.
Intra-cluster Ethernet traffic---including the telemetry stream itself---is carried over six Dell S5248F-ON switches, while out-of-band management relies on three Dell N1548 switches.
Two NVIDIA/Mellanox Quantum InfiniBand switches, an MQM8700 (HDR, 200~Gb/s) and an MQM9700 (NDR, 400~Gb/s), form the low-latency fabric used by the HPC and GPU partitions.
Finally, two dedicated gateway nodes perform perimeter routing and firewalling.
As these devices are closed appliances that cannot host the collection agent, their power draw is captured externally at the PDU level, consistent with the comprehensiveness requirement (\autoref{req:R4}); this is the only part of the infrastructure for which no node-local telemetry exists.

\end{document}